 \documentclass{article}

\usepackage{amsmath}
\usepackage{amsfonts}
\usepackage{graphicx}
\usepackage{caption}
\usepackage{subcaption}

 \usepackage{amssymb}

\usepackage{epstopdf}
\usepackage{epsfig}
\usepackage{hyperref}

\begin{document}
\title{A Higgs--Chern-Simons gravity 
model in $2+1$ dimensions}
\author{{\large }
{\large Eugen Radu}$^{\dagger}$
and {\large D. H. Tchrakian}$^{\ddagger \star}$ 
\\ \\
$^{\dagger}${\small Departamento de F\'isica da Universidade de Aveiro and CIDMA,
 Campus de Santiago, 3810-183 Aveiro, Portugal}
\\ 
$^{\ddagger}${\small School of Theoretical Physics, Dublin Institute for Advanced Studies,
10 Burlington Road, Dublin 4, Ireland}\\
$^{\star}${\small Department of Mathematical Physics, National University of Ireland Maynooth, Maynooth, Ireland}}

\date{}
\newcommand{\dd}{\mbox{d}}
\newcommand{\tr}{\mbox{tr}}
\newcommand{\la}{\lambda}
\newcommand{\La}{\Lambda}
\newcommand{\bt}{\beta}
\newcommand{\del}{\delta}
\newcommand{\ep}{\epsilon}
\newcommand{\ta}{\theta}
\newcommand{\ka}{\kappa}
\newcommand{\f}{\phi}
\newcommand{\vf}{\varphi}
\newcommand{\F}{\Phi}
\newcommand{\al}{\alpha}
\newcommand{\ga}{\gamma}
\newcommand{\de}{\delta}
\newcommand{\si}{\sigma}
\newcommand{\Si}{\Sigma}
\newcommand{\bnabla}{\mbox{\boldmath $\nabla$}}
\newcommand{\bomega}{\mbox{\boldmath $\omega$}}
\newcommand{\bOmega}{\mbox{\boldmath $\Omega$}}
\newcommand{\bsi}{\mbox{\boldmath $\sigma$}}
\newcommand{\bchi}{\mbox{\boldmath $\chi$}}
\newcommand{\bal}{\mbox{\boldmath $\alpha$}}
\newcommand{\bpsi}{\mbox{\boldmath $\psi$}}
\newcommand{\brho}{\mbox{\boldmath $\varrho$}}
\newcommand{\beps}{\mbox{\boldmath $\varepsilon$}}
\newcommand{\bxi}{\mbox{\boldmath $\xi$}}
\newcommand{\bbeta}{\mbox{\boldmath $\beta$}}
\newcommand{\ee}{\end{equation}}
\newcommand{\eea}{\end{eqnarray}}
\newcommand{\be}{\begin{equation}}
\newcommand{\bea}{\begin{eqnarray}}

\newcommand{\ii}{\mbox{i}}
\newcommand{\e}{\mbox{e}}
\newcommand{\pa}{\partial}
\newcommand{\Om}{\Omega}
\newcommand{\om}{\omega}
\newcommand{\vep}{\varepsilon}
\newcommand{\bfph}{{\bf \phi}}
\newcommand{\lm}{\lambda}
\def\theequation{\arabic{equation}}
\renewcommand{\thefootnote}{\fnsymbol{footnote}}
\newcommand{\re}[1]{(\ref{#1})}
\newcommand{\R}{{\rm I \hspace{-0.52ex} R}}
\newcommand{\N}{{\sf N\hspace*{-1.0ex}\rule{0.15ex}%
{1.3ex}\hspace*{1.0ex}}}
\newcommand{\Q}{{\sf Q\hspace*{-1.1ex}\rule{0.15ex}%
{1.5ex}\hspace*{1.1ex}}}
\newcommand{\C}{{\sf C\hspace*{-0.9ex}\rule{0.15ex}%
{1.3ex}\hspace*{0.9ex}}}
\newcommand{\eins}{1\hspace{-0.56ex}{\rm I}}
\renewcommand{\thefootnote}{\arabic{footnote}}

\maketitle


\bigskip

\begin{abstract}
  We study a gravity model in $2+1$ dimensions, arising from a generalized Chern-Simons (CS) density we call a 
	Higgs-Chern-Simons (HCS) density.
  This generalizes the construction of gravitational systems resulting from non-Abelian CS densities in all odd dimensions.
  The new HCS densities employed here are arrived at by the usual one-step descent of new Higgs--Chern-Pontryagin (HCP)
  densities, the latter resulting from the dimensional reduction of Chern-Pontryagin (CP) densities in some even dimension, 
	such that in any given dimension (including even) there is an infinite tower of such models. Here, we  
  restrict  our attention to the lowest dimension, $2+1$, and to the simplest such model resulting from the dimensional reduction of
  the $3$-rd CP density. We construct a black hole (BH) solution in closed form, generalizing the familiar BTZ BH. We also
  study the electrically charged BH solution of the same model augmented with a Maxwell term, and contrast this solution with
  the electrically charged BTZ BH, specifically concering their respective thermodynamic properties.
 
\end{abstract}
\medskip

\section{Introduction}
Chern-Simons gravities (CSG) derived from non-Abelian Chern-Simons (CS) densities in $2+1$ dimensions were proposed by Witten in
Ref.~\cite{Witten:1988hc} and they were extended to all odd dimensions by Chamseddine in Refs.~\cite{Chamseddine:1989nu,Chamseddine:1990gk}.
CSG models consist of superpositions of gravitational models of all possible higher order gravities
in the given dimensions, each appearing with a precise
real numerical coefficient resulting from the calculus. In this report we refer to these models, aka. Lovelock models,
as $p$-Einstein gravities, the number $p$ being the power of the Riemann curvature in the Lagrangian, with $p=0$ being the
cosmological constant.

The recent work \cite{Tchrakian:2017fdw} has proposed
 a new formulation  of the CSG systems, 
which allows their construction in all, $both$ odd and even dimensions. The derivation
of the new-CS densities follows exactly the same method as the $usual$-CS densities in odd dimensions. The $usual$ CS density results
from the one-step descent of the corresponding Chern-Pontryagin (CP) density.
The CP density being a total-divergence
\[
\Om_{\rm CP}=\pa_i\Om_i\ ,\quad i=\mu,D\ ;\ \quad\mu=1,2,\dots,d\ ;\ d=D-1\ ,
\]
the CS density is defined as the $D$-th component of $\Om_M$, namely $\Om_{\rm CS}\stackrel{\rm def.}=\Om_D$.

In the new formulation, 
the role of the usual-CP density, which is defined in even dimensions only, is played by what we refer to as the
Higgs--Chern-Pontryagin (HCP) density, described in Refs.~\cite{Tchrakian:2010ar,Radu:2011zy} and in Appendix A of Ref.~\cite{Tchrakian:2015pka}.
These are dimensional descendents of the $n$th CP density in $N=2n$ dimensions, down to residual $D$ dimensions ($D<N=2n$), where now $D$
can be either odd or even. The relics of the gauge connection on the co-dimension are Higgs scalars. The remarkable property of the
HCP density $\Om_{\rm HCP}[A,\F]$, which is now given in terms of both the residual gauge field $A$ and the Higgs scalar $\F$, is that like the CP density it
is also a $total\ divergence$
\[
\Om_{\rm HCP}=\pa_i\Om_i\ ,\quad i=\mu,D\ ;\ \quad\mu=1,2,\dots,d\ ;\ d=D-1\ .
\]

Implementing now the one-step descent of the density $\Om_i$, one can define the corresponding new Chern-Simons density
as the $D$-th component of $\Om_i$, namely $\Om_{\rm HCS}\stackrel{\rm def.}=\Om_D$.
We refer to the quantity $\Om_{\rm HCS}$ as the Higgs--Chern-Simons (HCS) density. A detailed definition of the HCS density is given in
Refs.~\cite{Tchrakian:2010ar,Radu:2011zy,Tchrakian:2015pka}. Subsequently, a similar definition for the HCS density was given 
in \cite{Szabo:2014zua},
but only in odd dimensions~\footnote{Apart from the definition in Ref.~\cite{Szabo:2014zua} being restricted to odd dimensions, there is another important difference with with our formulation. In our case the
dimensional reduction is carried out on the {\bf gauge\ invariant} CP density yielding the HCP density, while in \cite{Szabo:2014zua} it is the CS density in the (higher) odd dimensions that is
subjected to dimensional reduction. While the results happen to be similar, there is no guarantee they should agree since subjecting a {\bf gauge\ variant} CS density to symmetry imposition as  done in \cite{Szabo:2014zua} problematic.} and with the Higgs scalar being a complex column, not suited to the application here. In our formulation, HCS densities are given in both odd and even dimensions and in any given dimension,
there is an infinite family of HCP densities in $D$ dimensions, as also HCS densities in $d=D-1$ dimensions, since these follow from the descent of a
CP density in any dimension $N=2n>D$.

Once the Higgs-CS (HCS) densities are calculated, they can be employed to generate gravitational theories in the same spirit
as in \cite{Witten:1988hc,Chamseddine:1989nu,Chamseddine:1990gk}, which could be designated as HCS gravities (HCSG).
In any given dimension, there is an infinite family of such theories, each resulting from the infinite family of HCS densities.

In the passage of the HCS densities to gravitational systems
in $d=D-1$ dimensions, we see from \re{oddd} and \re{higgsevenD2} below, that the gauge group is
chosen to be $SO(d)$ and the Higgs multiplet is chosen to be a
$D$-component $isovector$ of $SO(D)$~\footnote{These choices coincide
with the representations that yield monopoles on $\R^d$ described in\cite{Tchrakian:2010ar} }.
The cornerstone of constructing CSG models is the identification of the non-Abelian (nA) $SO(D)$ connection in $d=D-1$
dimensions~\footnote{We do not make a choice for the signature of the space at this stage.},
with the spin-connection $\om_\mu^{ab}$ and the $Vielbein$ $e_\mu^a$, ($\mu=1,2,3$;\ a=1,2,3). The
prescription employed in \cite{Witten:1988hc,Chamseddine:1989nu,Chamseddine:1990gk} is
\bea
A_{\mu}&=&-\frac12\,\om_{\mu}^{ab}\,\ga_{ab}+\ka\,e_{\mu}^{a}\,\ga_{a4}\quad\Rightarrow\quad
F_{\mu\nu}=-\frac12\left(R_{\mu\nu}^{ab}-\ka^2\,e_{[\mu}^{a}\,e_{\nu]}^{b}\right)\ga_{ab}+\ka\,C_{\mu\nu}^{a}\ga_{a4}\,,\label{oddd}
\eea
$(\ga^{ab},\ga^{a4})$ being the Dirac gamma matrices used in the representation of the algebra of $SO(D)\,,\ D=4$. The constant $\ka$ has
dimensions $L^{-1}$, compensating the difference of the dimensions of the spin-connection and the $Dreibein$. In \re{oddd},
$$R_{\mu\nu}^{ab}=\pa_{[\mu}\om_{\nu]}^{ab}+(\om_{[\mu}\om_{\nu]})^{ab}$$ is the Riemann curvature, and $C_{\mu\nu}^a=D_{[\mu}e_{\nu]}^a$ is the torsion.

Here, in addition to \re{oddd}, we posit the corresponding prescription for the Higgs scalar $\F$,
\bea
2^{-1}\F=(\f^{a}\,\ga_{a5}+\f\,\ga_{45})\  &\Rightarrow& \ 
2^{-1}D_{\mu}\F=(D_{\mu}\f^{a}-\ka\,e_\mu^a\,\f)\ga_{a5}+(\pa_{\mu}\f+\ka\,e_\mu^a\,\f^a)\ga_{45}
\label{higgsevenD2}
\eea
which clearly displays the $iso$-$four$-$vector$ $(\f^a,\f^4)$, that are split into the $three\ component$ frame-vector
field $\f^a$ and the scalar field $\f=\f^4$. The covariant derivative $D_\mu\F$ of the Higgs scalar features the gravitational covariant
derivative $$D_\mu\f^a=\pa_\mu\f^a+\om_\mu^{ab}\f^b$$ of the frame vector field $\f^a$.  
 Indeed, this is a vector field
$\f_\mu=e_\mu^a\f^a$, which however has rather unusual dynamics as will be seen below. It is neither a gauge field nor a
Proca field, rather, it has geometric content. 

The fields $(\f^a,\f)$ are not matter fields like, $e.g.$ Maxwell or Yang-Mills, or Skyrme, $etc$. In the latter
cases, the covariant derivatives are not defined by the (gravitational) spin-connection, while here they are as
seen in \re{higgsevenD2}.   In this sense they are like spinor fields. 
 An immediate consequence
of this is that theories like the one proposed here can support solutions with torsion.

  Unlike spinors however,  the fields $(\f^a,\f)$ are gravitational coordinates as they originate from the Higgs field
$\F$ of the nA gauge theory, which
itself is a (dimensional) descendent of a (higher dimensional) connection. Thus, as seen from \re{higgsevenD2}, $(\f^a,\f)$
are on the same footing as the $usual$ gravitational coordinates $(\om_\mu^{ab},e_\mu^a)$. 
As a consequence, we would expect that the effect of $(\f^a,\f)$ on the solutions cannot be characterised as $hair$. We
expect that they support only  black hole solutions and do not describe horizonless (soliton)
solutions in the limit of the horizon radius vanishing, as it happens in the usual theories with hair.

 The gravitational models resulting from the Higgs-CS (HCS) densities via \re{oddd}-\re{higgsevenD2} are
referred to as HCS gravities~\cite{Tchrakian:2017fdw} (HCSG). In this report we have restricted our detailed study to the
``simplest'' HCSG model in $2+1$ dimensions, namely to the HCSG model resulting from the HCS density descended from
the HCP density in $6$ (rather than those in $8$, $10$, $etc.$) dimensions. This model is an extension of the
$usual$ Chern-Simons gravity~\cite{Witten:1988hc}, and our solutions can be contrasted with the 
Banados, Teitelboim and Zanelli 
(BTZ)~\cite{Banados:1992wn}
black hole solution in that theory. 
 
This paper is organized as follows.
The model studied is presented in Section {\bf 2}, including the equations of motion. Then in Section {\bf 3} the imposition of (radial)
symmetry is carried out, for the torsion-free case where the spin-connection is restricted to the Levi-Civita connection. There are two
questions that symmetry imposition must address: $a$) The consistency of the Ansatz, and $b$) the consistency of using a torsion-free Ansatz,
in a theory where the torsion tensor cannot be set to zero $a\ priori$ because of the appearance of the gravitational covariant derivative.
Both these questions are addressed implicitly, in Section {\bf 3}. In Section {\bf 4} we present the solutions of the
system  with backreacting HCSG fields.

\section{A HCSG model in $2+1$ dimensions}
The Higgs--Chern-Simons density (HCS) employed here is the ``simplest'' example in $2+1$ dimensions.
By simplest we mean that the Higgs--Chern-Simons (HCS) density employed to construct the HCS gravity (HCSG), is the one resulting from the
``simplest'' Higgs--Chern-Pontryagin (HCP) density, which is defined in one dimension higher, namely in $four$ dimensions. Now in $four$ dimensions,
HCP densities can be constructed as dimensional descendants of a CP density in $2n>4$ dimensions, hence it is reasonable to describe the
``simplest'' case at hand to be the HCP density in $4$ dimensions, the one that descends from the CP density in $2n=6$ dimensions.

Since like the CP density, the HCP density is a $total\ divergence$, then the corresponding HCS density results from usual $one$-$step$ descent,
in this case from $4$ to $3$ dimensions. It may be helpful to display two such HCS densities in $2+1$ dimensions, each resulting from the
$one$-$step$ descent of a HCP density th $4$ dimensions, the first of which has resulted from the dimensional descent of the CP in $6$
dimensions, and the second in $8$ dimensions. These are
\bea
\Omega^{(3,6)}_{\rm HCS}&=&
-2\eta^2\Omega_{\rm CS}^{(3)}-\vep^{\mu\nu\la}\mbox{Tr}\,\ga_5D_{\la}\F\left(F_{\mu\nu}\,\F+F_{\mu\nu} \F\right)\,
,
\label{HCS36}
\\
\Omega^{(3,8)}_{\rm HCS}&=&6\eta^4\Omega_{\rm CS}^{(3)}-\vep^{\mu\nu\la}\,\mbox{Tr}\,\ga_5\,\bigg\{
6\,\eta^2\left(\F\,D_{\la}\F-D_{\la}\F\,\F\right)\,F_{\mu\nu}
\nonumber\\
&&\hspace{20mm}-\bigg[\left(\F^2\,D_{\la}\F\,\F-\F\,D_{\la}\F\,\F^2\right)-2\left(\F^3\,D_{\la}\F-D_{\la}\F\,\F^3\right)\bigg]F_{\mu\nu}
\bigg\}\,,\label{HCS38}
\eea
where the leading term $\Om_{\rm CS}^{(3)}$ in each is the usual CS density
\be
\label{CS3}
\Om_{\rm CS}^{(3)}=\vep^{\mu\nu\la}\mbox{Tr}\,A_{\la}\left(F_{\mu\nu}
-\frac23\,A_{\mu}A_{\nu}\right),
\ee
and where the constant $\eta$ and the Higgs field $\F$ both have the dimensions of $L^{-1}$, like the gauge connection.

Our choice for the ``simplest'' HCS density will be \re{HCS36}, rather than for example \re{HCS38} or ones originating from even higher
dimensional CP densities. Applying the prescriptions \re{oddd} and \re{higgsevenD2}, \re{HCS36} yields the HCS gravitational (HCSG) model
studied here, 
\bea
{\cal L}_{\rm HCSG}&=&\vep^{\la\mu\nu}\vep_{abc}\Bigg\{2\eta^2\ka\,\left(e_{\la}^{c}\,R_{\mu\nu}^{ab}
-\frac23\ka^2e_{\mu}^{a}e_{\nu}^{b}e_{\la}^{c}\right)\nonumber\\
&&+\bigg[2(R_{\mu\nu}^{ab}-\ka^2\,e_{[\mu}^{a}e_{\nu]}^{b})\left[\f^{c}(\pa_{\la}\f+\ka\,e_{\la}^d\f^d)
-\f (D_{\la}\f^{c}-\ka\, e_{\la}^c\f)\right]\nonumber\\
&&\qquad\qquad\qquad\qquad\qquad\qquad-4\ka\f^{a}(D_{\la}\f^{b}-\ka\,e_\la^b\f)C_{\mu\nu}^{c}\bigg]\Bigg\},
\label{HCSG2+1}
\eea
where $C_{\mu\nu}^{c}$ is the torsion tensor.

In the Lagrangian \re{HCSG2+1}, both the torsion $C_{\mu\nu}^{c}$ and the covariant derivative $D_{\la}\f^{c}$ are defined
in terms of the spin-connection $\om_\mu^{ab}$, so that the variation w.r.t. to the latter will result in the appearance of the
$C_{\mu\nu}^{c}$ in the field equations. Whether or not can the torsion be set equal to zero
consistently~\footnote{Note that coupling gravity, $e.g.$ to Maxwell, Yang-Mills or Skyrme/Higgs
fields can consistently be considered in a torsionless framework, since the kinetic terms of these fields $do\ not$ feature the spin-connection.}
 for a particular field configuration must be checked. This question is tackled in Section {\bf 3}
where the system is subjected to static radial symmetry.

In the present work, we restrict our attentions to the Levi-Civita connection
\be
\label{L-C}
\om_{\mu}^{ab}[e,\pa e]=-\frac12\,e_{\mu}^d\,e^{\rho a}e^{\si b}\pa_{[\rho}e_{\si] d}+\frac12\,e^{\la [a}\,\pa_{[\mu}e_{\la]}^{b]}\,,
\ee
subject to checking the consistency of our Ansatz with this case.

\subsection{Equations of motion}
To express the equations of motion concisely, it is useful to introduce the notations
\bea
{\cal V}^{\la}_c&=&\vep^{\la\mu\nu}\vep_{cab}(R_{\mu\nu}^{ab}-\ka^2e_{[\mu}^ae_{\nu]}^b)\,,\label{calV}\\
{\cal S}&=&\eta^2-\f^2-|\vec\f|^2\label{calS}\,.
\eea
The (modified) Einstein equations 
follow from the variation of \re{HCSG2+1}
$w.r.t.$
 the Dreibein $e_\la^c$, yielding
\bea
\label{Ein}
E_c^\la&=&\ka\,{\cal S}{\cal V}^{\la}_c+4\ka\vep^{\la\mu\nu}\vep_{cab}\left[D_{\mu}\f^aD_{\nu}\f^b
-\ka\f\left(2D_\mu\f^a-\ka\,e_\mu^a\,\f\right)e_\nu^b\right]=0\,.
\eea
The equations 
resulting from the variation of the frame-vector scalar $\f^c$ are
\be
\label{fa}
E_c=2{\cal V}^{\la}_c(\pa_\tau\f+\ka\ \vec e_\la\cdot\vec\f)+4\vep^{\la\mu\nu}\vep_{cab}\,C_{\mu\nu}^a\,(D_\la\f^b-\ka\,e_\la^b\f)=0\,,
\ee
while the equation
resulting from the variation of the scalar $\f$ is
\be
\label{eqf}
E=-2{\cal V}^{\la}_c\,(D_\tau\f^c-\ka\,e_\la^c\f)=0\,.
\ee
The torsion equations 
follow from the variation of \re{HCSG2+1} w.r.t. the spin connection $\om_\la^{ab}$,
not constrained to be the Levi-Civita connection \re{L-C}, yielding
\bea
E_{ab}^\la&=&\vep^{\la\mu\nu}\vep_{cab}\Bigg\{-2\ka\,{\cal S}C_{\mu\nu}^c+8D_\mu\f^c\,\pa_\nu\f+\nonumber\\
&&\qquad\qquad+8\ka\left[(\vec e_\nu\cdot \vec\f)D_\mu\f^c+e_\nu^a\,\f\pa_\mu\f\right]-8\ka^2\f(\vec\f\cdot\vec e_\nu)e_\mu^c\Bigg\}=0
\label{torseq}
\eea
Note that in \re{torseq} the curvature $R_{\mu\nu}^{ab}$ does not appear. Setting $C_{\mu\nu}^a=0$
in \re{torseq} puts a constraint on the fields $(\f^a,\f)$,
which for consistency should be satisfied for a given field configuration.

In \re{Ein}, in \re{torseq} and in \re{fa}, the ``usual'' notation $|\vec\f|^2=\f^a\f^a$, $(\vec e_\tau\cdot\vec\f)
=e_\mu^a\f^a$ and $(\vec e_\mu\cdot D_\nu\vec\f)=e_\mu^aD_{\nu}\f^a$ is used.

Also, in what follows we opt for Minkowskian signature and make the replacement
\be
\label{replace}
\ka\to i\ka\ ,\quad h\to -ih\,.
\ee
As such, setting $\phi=\phi^a=0$
results in Einstein gravity with a negative cosmological constant
$\Lambda=-\kappa^2$.
As found in \cite{Banados:1992wn}
by  Banados, Teitelboim and Zanelli 
(BTZ),
this model possess black hole (BH) solutions with
Anti-de Sitter asymptotics.
Their study has been seminal for a better understanding of
BH physics and dualities.

A natural question, which we propose to address in the
following Section,
is if the BTZ BHs possess generalizations
within the full model
(\ref{HCSG2+1})
with excited functions $\phi,\phi^a$.

\section{The solutions}

\subsection{The Ansatz}

We convert the usual static radially symmetric metric Ansatz in $2+1$ dimensions
\bea
\label{metric}
ds^2=A^2(r)dr^2+r^2d\vf^2-B^2(r)dt^2,
\eea
(where
$r,t$ are the radial and time coordinate, respectively,
while $0\leq \varphi <2\pi$),
to the corresponding Dreibein Ansatz as follows
\bea
\label{DreiAnsatz}
&&e_{r}^{\al}=A\,n_{(1)}^{\al}\ ,\qquad e_{\vf}^{\al}=r\,n_{(2)}^{\al}\ ,\qquad e_{t}^{3}=B,
\label{e+}
\eea
where
\be
\label{n}
n_{(1)}^{\al}=\left(\begin{array}{l}
\cos n\vf\\
\sin n\vf
\end{array}
\right)=(\vep n_{(2)})^\al\ ,\quad
n_{(2)}^{\al}=\left(\begin{array}{l}
-\sin n\vf\\
\ \ \cos n\vf
\end{array}
\right)=-(\vep n_{(1)})^\al,
\ee
where $n$ is an integer.

The Ansatz we use for the frame-vector field $\f^a=(\f^{\al},\f^3)$ and the scalar $\f$ is
\bea
\f^{\al}&=&f(r)\,n_{(1)}^{\al}\ ,\quad\f^3=g(r)\label{fal3},
\\
\f&=&h(r)\,,
\label{f}
\eea
having used the same unit vector $n^{\al}_{(1)}$ in \re{fal3} as that given by \re{n}.
 
We restrict our attention to the torsion-free case, substituting the Ansatz \re{e+} 
in the Levi-Civita connection
\re{L-C} and calculating the resulting components of the curvature. Using this reduced spin connection and the Ansatz
\re{fal3}-\re{f}, we calculate the reduced (covariant) derivatives of the fields $(\f^a,\f)$. 
Substituting all this in the
equations \re{Ein}, \re{fa} and \re{eqf}, we find that these are satisfied for $g(r)=0$. 
Thus there remain
four equations for the functions $(A,B,f,h)$, plus one constraint equation that is satisfied.
We then substitute this torsion-free Ansatz in the torsion equations \re{torseq}, yielding three non-trivial equations,
each of which is satisfied by the ``Einstein'' equation for $h'$.
Therefore 
we conclude that our torsion-free Ansatz is consistent.

We also remark that the system is described by the 
reduced Lagrangian resulting from the imposition of symmetry on \re{HCSG2+1},
\bea
\label{redlag}
4^{-1}L_{\rm HCSG}&=&2\eta^2\ka\left[-\frac{BA'}{A^2}+\ka^2rAB\right]\nonumber\\
&&-\bigg\{2\left(\frac{B'}{A^2}\right)fh'+\frac{\ka}{A}\left[\left(\frac{BA'}{A}\right)(g^2-h^2)-B'(f^2+2rhh')\right]\nonumber\\
&&\qquad+\ka^2\left[(rB'+B)fh-rB(fh'-hf')\right]+\ka^3rAB(f^2+g^2-3h^2)\bigg\}\,.
\eea 
We observe that, indeed, \re{redlag} does not feature the derivative of the function $g(r)$, and hence one can
consistently set $g(r)=0$.

   \subsection{An exact solution (HBTZ BH)
   }
  In the absence of the fields $(\f^a,\f)$, only the equation \re{Ein} must be solved, which is of course satisfied
by the vacuum BTZ solution~\cite{Banados:1992wn}. Since the BTZ solution satisfies the equations of
Chern-Simons gravity
(CSG) in $2+1$ dimensions, and since the solution presented here results instead from the equations of the Higgs--Chern-Simons
gravity (HCSG), one might refer to the new solutions as Higgs-BTZ, or simply as HBTZ black holes. 

We have found that the field equations  (\ref{Ein})-(\ref{eqf}) possess a closed-form solution with
\bea
B^2(r)&=&\frac{1}{A^2(r)}=\ka^2  r^2+c_t,
\label{solnsAB}
\\
h(r)&=&c_0r\ ,\quad f(r)=\frac{c_0}{\ka}B(r),
\label{solnshf}
\eea
where $c_t, c_0$ are free parameters. Equivalently, \re{solnshf} can be replaced by $(f,h)\to(h,f).$

One can see that (\ref{solnsAB})
corresponds to a globally AdS$_3$
geometry (for $c_t>0$) or to a BTZ black hole (for $c_t<0)$,
which are solutions of  the Einstein equations also for $\phi^a=\phi=0$.
Thus the fields $(\phi^a,\phi)$ 
do not backreact on these geometries, leading to
a vanishing {\it effective} energy-momentum tensor, $i.e.$ that these solutions are {\it effectively} vacuum solutions~\footnote{
   It is interesting to note the analogy with  gravitating  self-dual instantonic Yang-Mills configurations
   with Euclidean metric 
see e.g. \cite{Charap:1977ww}, \cite{Brihaye:2006bk}, \cite{Radu:2007az}.  In that case the reason is that the stress tensor vanishes identically due to self-dualty. In the present case, (\ref{solnsAB}),  (\ref{solnshf})
are solutions also for an Euclidean signature 
and one can easily check that the contribution of the 
second term in (\ref{HCSG2+1}) 
to the total action is nonvanishing. 
Based on this analogy, one might claim that these closed form solutions are {\it effectively} vacuum solutions.
}.
This solution possesses another unusual property, namely that $f$ and $h$ functions diverge at infinity.

One may ask if other solutions exist  
apart from  (\ref{solnsAB}),  (\ref{solnshf}) 
(preferably with both $f$ and $g$
finite in the far field).
The answer seems to be negative,
a strong indication
in this direction coming from the study of the near horizon expansion of the solutions.
The first step here is to notice that the 
field equations imply that the function $f$ can be eliminated, with
 \begin{eqnarray}
f(r)=\frac{h'(r)}{\kappa A(r)},
\end{eqnarray}
The BH solutions
possess an  horizon at $r=r_H>0$,
where we suppose  the following approximate form of the 
generic solution
(with 
$A(r)=\frac{1}{\sqrt{U(r)}},~B(r)=\sqrt{H(r)}$):
   \begin{eqnarray}
     \label{pert}
H(r)=\sum_{k\geq 0}H_{(k)}(r-r_H)^k,~~
U(r)=\sum_{k\geq 0}U_{(k)}(r-r_H)^k,~~
h(r)=\sum_{k\geq 0}h_{(k)}(r-r_H)^k,~~
\end{eqnarray}
where $H_{(k)}$, $U_{(k)}$ and $h_{(k)}$ are real numbers.
After substituting  \re{pert} in the equations of motion and solving order by order, we have found 
that 
\begin{eqnarray} 
&&
H(r)= H_{(1)}(r-r_H)+\frac{H_{(1)}}{2r_h}(r-r_H)^2+O(r-r_H)^{12},
\\
&&
U(r)= 2\kappa^2 r_H(r-r_H)+\kappa^2 (r-r_H)^2+ O(r-r_H)^{12},
\\
&&
h(r)=\frac{h_{(0)}}{r_H}r+O(r-r_H)^{11}.
\end{eqnarray}
The above result has been proven up to order $11$ in
perturbation theory. It is likely however, that it holds to all orders, although we do not have a proof of that.
This coincides with the near horizon form of the exact solution
(\ref{solnsAB}),
(\ref{solnshf})  
(note that the metric function $B(r)$ is fixed up to a constant factor 
$B(r)\to \lambda B(r)$ ).
Thus we conclude that
(\ref{solnsAB}),
(\ref{solnshf})  
is
likely 
 the unique configuration compatible with a regular expansion at the horizon\footnote{This also 
agrees with our numerical results, which
have failed to indicate the existence of other solutions apart from 
(\ref{solnsAB}),
(\ref{solnshf}).}.

A similar argument excludes the existence of particle-like solitonic solutions
apart from  
(\ref{solnshf}) 
in an  AdS background.

\subsection{Deformed charged HBTZ black holes}

The above results,  namely the HBTZ solution, are 
consistent with the spirit of the 'no hair' theorems, which
exclude the existence of BH solutions with matter fields that do not possess 
measured quantities subject to  a Gauss Law
\cite{Bekenstein:1996pn},
\cite{Volkov:1998cc},
\cite{Herdeiro:2015waa}.

The 'no hair' constraints can be circumvented for
more complicated models,
typically possessing gauge fields
(see $e.g.$
\cite{Gubser:2008px},
\cite{Gubser:2008zu},
\cite{Winstanley:1998sn},
	for seminal work on hairy black holes with AdS asymptotics).
Thus one can ask if the results in the previous subsection are generic and hold also in more general models,
with a Lagrangian containing matter
fields in addition to fields  $(e_\mu^a,\phi^a,\phi)$.

To address this question, we consider the simplest generalization of the model 
(\ref{HCS38}), 
with an additional Maxwell term,
\begin{eqnarray}
\label{new-model}
{\cal L} = {\cal L}_{\rm HCSG}-\frac{1}{4}F_{\mu\nu}^2,
\end{eqnarray}
with $F_{\mu\nu}=\pa_{[\mu}A_{\nu]}$ the field strength tensor. It is clear that in this case there will be a
nonvanishing stress tensor. 

Considering the same {\bf metric} Ansatz (\ref{metric}),
we take a purely electric connection
\begin{eqnarray}
A_\mu=(A_0,A_i)=(V(r), 0)~.
\end{eqnarray}
The corresponding equations are easily derived;
note that 
$V(r)$ interacts with $(\phi^a, \phi)$ only via the geometry, with the existence of a first integral
\begin{eqnarray}
     \label{firstint}
V=\int dr \frac{Q}{r}\sqrt{AB},
\end{eqnarray}
where $Q$ is an integration constant
identified with the electric charge.

For $f=g=h=0$,
the  electrically  charged BTZ BH \cite{Banados:1992wn} is recovered, with
\be
\label{charged-BTZ}
B^2(r)=\frac{1}{A^2(r)}=\ka^2r^2-M-\frac{Q^2}{2\kappa}\log r~.
\ee
This solution possess an event horizon at $r=r_H$, where 
$B(r_H)=0$.

We are interested in generalizations of this solution with nonvanishing fields $(f,h)$.
Assuming the existence of a regular event horizon at $r=r_H>0$, the field equations
imply the following approximate form of the solutions near the horizon,
\begin{eqnarray}
\label{eh1}
&&
\frac{1}{A^2(r)}=H_1 \kappa^2 r_H^2 (r-r_H)+\dots,~~
B^2(r)=\kappa^2 r_H^2 u_1(r-r_H)+\dots,~~
\\
\nonumber
&&
f^2(r)=h_1^2 H_1r_H^2(r-r_H)+\dots,~~
h(r)=h_0+h_1(r-r_H)+\dots,~~
\end{eqnarray} 
in terms of two free parameters $(h_0,u_1)$, with
\begin{eqnarray}
\label{eh2}
&&
H_1=\frac{2}{r_H}
\left(
1+\frac{Q^2}{6(2+h_0^2)\kappa^3 r_H^2}
 (
\sqrt{1-\frac{12h_0^2\kappa^3 r_H^2}{Q^2}}-1
 )
\right),
\\
&&
\nonumber
h_1=\frac{1}{2h_0H_1r_H^2}
\left(
(2+3h_0^2)(2-H_1r_H)+2h_0^2H_1r_H
-\frac{Q^2}{\kappa^3r_H^2}
\right)~.
\end{eqnarray}
Note that the condition for $H_1$ to be real
implies the existence of a maximal value of the functions $h(r)$
at the horizon
\begin{eqnarray}
\label{cond}
h(r_H)<\frac{Q}{2\kappa r_H \sqrt{3\kappa}}\,.
\end{eqnarray}

 \begin{figure}[h!]
\begin{center}
\includegraphics[width=0.55\textwidth]{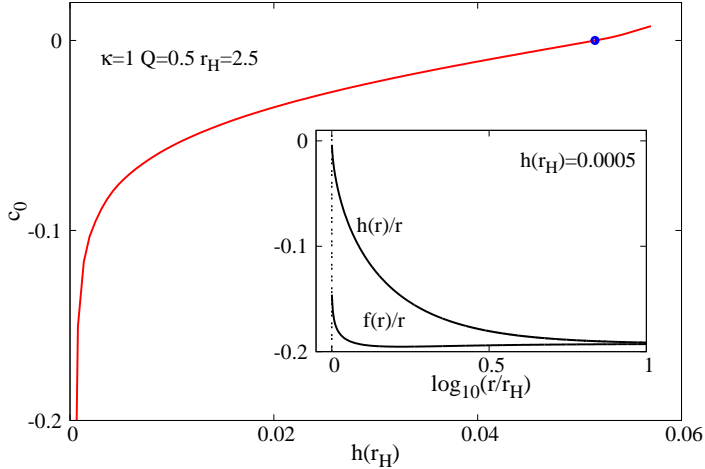} 
\caption{ The parameter $c_0$
which enters the far field asymptotics of the functions $f(r),h(r)$
(with  $(h(r),f(r)) \to c_0 r$),
is shown as a function of the value of $h(r)$ at the horizon,
for as set of generic solutions.
The inset shows a typical profile of the functions $f(r),h(r)$.
}
\label{datagen}
\end{center}
\end{figure} 

  We seek electrically charged HBTZ solutions deforming the charged BTZ BH's, with the functions $(f,h)$ excited. We have encountered solutions that possess isolated zeros of $(f,h)$ for some values of $r$, $i.e.$, displaying nodes, whose detailed study we have eschwewd. Henceforth we concentrate exclusively on {\it nodeless, fundamental} solutions. 
   
Generic solutions with non-standard asymptotics are found for the near horizon expansion (\ref{eh1}).
As seen in Figure \ref{datagen},
similar to the exact solution
(\ref{solnshf}),
the fields $(f,h)$ diverge linearly as $r\to \infty$, 
with  $(h(r),f(r)) \to c_0 r$.
This divergence mixes with the logarithmic terms originating in the Maxwell
part of the theory, leading to a slower decay at infinity of the functions $B^2/r^2$ and $A^2r^2$, as compared to the charged BTZ case.

 In addition to these generic solutions, we have found special solutions isolated in parameter space. 
For given (and nonzero) $(r_H,Q)$, a different situation is found for a particular set of 
near-horizon parameters $(h_0,u_1)$, which lead to $f$ and $h$ vanishing at infinity, with $c_0=0$
(for the data exhibited in Figure \ref{datagen}, this corresponds to the blue point).

The resulting configurations possess the following large-$r$ asymptotic expansion: 
\begin{eqnarray}
\label{inf1}
&&
 A^2(r)=\frac{1}{\kappa^2 r^2}+
\left(
\frac{c_t-4  c_s^2}{\kappa^2}+\frac{ Q^2}{2\kappa^5}\log r
\right)
\frac{1}{r^4}
+\dots,
~~
B^2(r)=
\kappa^2 r^2-c_t  -\frac{  Q^2}{2\kappa^5}\log r+\dots,
\\
&&
f(r)=-\frac{c_s}{r}+\dots,~\qquad\qquad\qquad\qquad\qquad\qquad\qquad h(r)=\frac{c_s}{r}+\dots,~
\end{eqnarray}
in terms of two constants
$c_t$,
$c_s$.
 
This set of  solutions are of particular interest and, for the remainder of
this Section we shall {\bf confine our} discuss{\bf ions to} their basic properties. 
These BHs are constructed numerically by using a standard ordinary differential equation solver.
In our approach, we evaluate
the initial conditions 
(\ref{eh1})
at $r = 10^{-5}$ for global tolerance $10^{-14}$, 
adjusting for fixed shooting
parameter and integrate the equations towards $r\to \infty$
where the far field asymptotics (\ref{inf1}) are approached.
The profiles of a typical solution is displayed in Figure 1.

 \begin{figure}[h!]
\begin{center}
\includegraphics[width=0.55\textwidth]{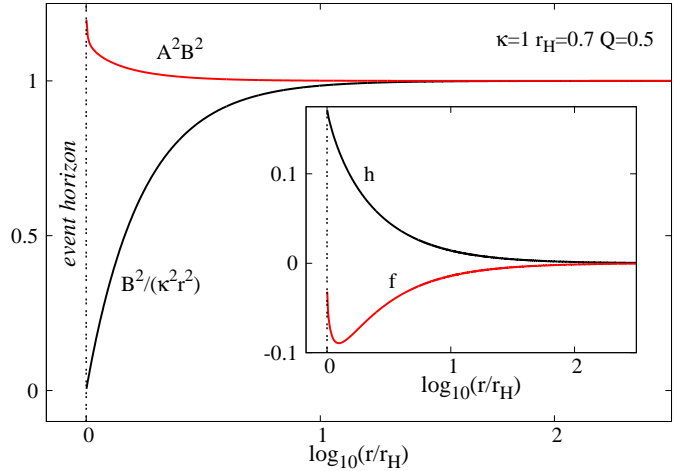} 
\caption{ The profile of a typical black hole solution with nontrivial
$(\phi^a,\phi)$-fields.
}
\label{profile}
\end{center}
\end{figure} 

 \begin{figure}[h!]
\begin{center}
\includegraphics[width=0.55\textwidth]{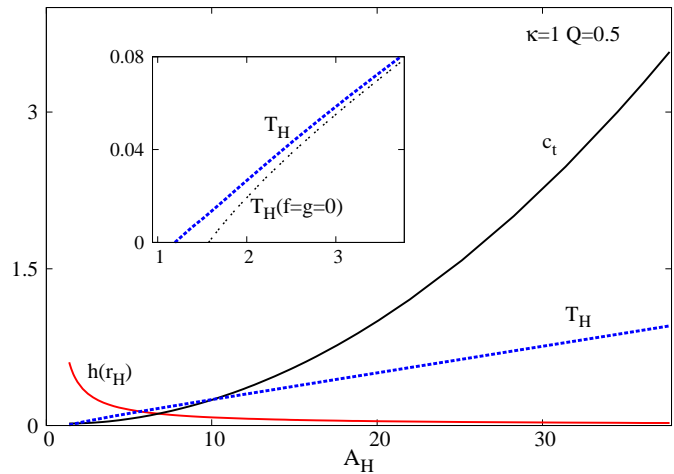} 
\caption{ Relevant data is shown for a family of charged black hole solutions.
}
\label{data}
\end{center}
\end{figure} 
 
   The resulting solutions can be interpreted as deformations of the charged BTZ BHs,
 since the fields $(\phi^a,\phi)$ in this case deform the geometry  
on and outside the horizon.
 To highlight the departure of the charged (deformed) HBTZ BH from the charged BTZ BH, we consider some relevant thermodynamic quantities. 
Their Hawking temperature $T_H$ and
event
 horizon area
$A_H$
 are unambiguously defined,
with
%
\begin{eqnarray}
\label{TH}
T_H=\frac{\kappa^2 r_H^2}{4\pi}\sqrt{H_1 u_1},~~A_H=2\pi r_H.
\end{eqnarray} 
For $f=h=0$, the mass of the solutions is determined by the constant $c_t$
in the far field asymptotics (\ref{inf1}),
while the entropy is $S=A_H/4$.
However,
given the direct coupling of the
 coupling of matter fields $\phi^a,\phi$ with the curvature tensor,
 the definition of the BH mass and entropy for the general model (\ref{new-model})
is not $a\ priori$ clear, this issue requiring a separate study.

A number of basic properties of the  solutions are shown in Figure 3.
One can see that, similar to the charged BTZ case, there is a single branch of solutions,
with both $c_t$ and $T_H$ increasing with $A_H$. 
Moreover, for the same horizon size, the solutions with nonzero $(f,h)$ are warmer.
Interestingly, the solutions possess a zero temperature extremal limit,
which is nonsingular and featurs a nonvanishing horizon size.
The deviation\footnote{The solutions possess a second parameter, $c_s$, in
the far field expansion (\ref{inf1}),
which can be interpreted as a scalar 'charge',
and provides annother measure for
the deviation of  charged HBTZ BH from the (usual) charged BTZ BH.
Note however, that no global charge is a associated with $c_s$.
} from the standard charged BTZ BH is maximal, close
to that limit,  while the fields $(\phi^a,\phi)$
trivialize for large horizon size,
$(f,g)\to 0$ as $r_H\to \infty$.

\section{Summary and outlook}
 
   In this paper we have studied a generalization of the familiar Chern-Simons gravity (CSG) in $2+1$ dimensions proposed in
   Ref.~\cite{Witten:1988hc}, where
   the gravitational model is described by both the Einstein-Hilbert (EH) Lagrangian with a (negative) cosmological constant. In our
   generalized model, in addition to the EH and cosmological terms, our Lagrangian features new terms described by a frame-vector
   field $\f^a$ and a scalar field $\f$. The dynamical terms of the fields $(\f^a,\f)$ are non-standard~\footnote{By standard, we mean
   that the dynamical term of the field $\f_\mu=e^a_\mu\f_a$ would feature the {\it square} of the velocity field $\pa_\mu\f_\nu$.}.

   The generalization that introduces the fields $(\f^a,\f)$ is a result of the following. 
	While the construction of Chern-Simons gravities
   in all odd dimensions employs~\cite{Witten:1988hc}\cite{Chamseddine:1989nu,Chamseddine:1990gk} non-Abelian Chern-Simons densities, we
   employ instead an alternative type of generaized CS densities that we have referred to as
   Higgs--Chern-Simons~\cite{Tchrakian:2010ar,Radu:2011zy,Tchrakian:2015pka} (HCS) densities. The HCS densities are extracted from the
   Higgs--Chern-Pontryagin~\cite{Tchrakian:2010ar} (HCP) densities, which being descendents of Chern-Pontryagin (CP) constitute an infinite
   family in any $even$ given, including {\it even} spacetime dimensions. 
This construction is described in Ref.~\cite{Tchrakian:2017fdw}.
The present work is a first exploration of such gravitational models, and we have chosen the lowest
   dimension, $2+1$, and simplest model in this dimension, which employs the HCS density extracted from the $3$-rd
   CP density in $6$ dimensions.

   The HCS gravity model we have studied, \re{HCSG2+1}, features the gravitational covariant derivatives of the frame vector field $\f^a$,
   $a=1,,2,3$ in this case, so that the possibility of finding non-zero torsion solutions is not excluded. Here we have chosen to seek only
   torsion-free solutions. We have constructed radially symmetric solutions employing the usual metric~\cite{Banados:1992wn} Ansatz,
   augmented by a suitable Ansatz for the fields $(\f^a,\f)$. We have verified the consistency of our Ansatz, and what is more is, that we
   have verified that the torsion equations resulting from the variation of \re{HCSG2+1} with respect to the spin-connection are identically
   satisfied by the equations resulting from the variation of \re{HCSG2+1} with respect to the metric fields, in {\it this} Ansatz.

   We have found a closed form solution analogous to the BTZ~\cite{Banados:1992wn} black hole, which we have referred to as a
   Higgs-BTZ HBTZ black hole (BH). We have a heuristic verification of the fact that this HBTZ BH is unique, and most importantly that
   there are no regular solutions in this model. In this respect, the HBTZ BHs cannot be considered as hairy solutions,
   as would have been the case if matter fields were present and the solutions persisted in the limit of vanishing
   horizon radius. As such, the fields $(\f^a,\f)$ should not be seen as matter fields.

   With the intention of introducing a matter field, we have extended our model to feature a Maxwell term.
   We have sought, and found, new solutions by adding the Maxwell field to \re{HCSG2+1}. As in the case of the electrically charged BTZ
   BH~\cite{Banados:1992wn}, the Abelian matter does not result in hairy solutions and the solutions we find are qualitatively similar to the
   electrically charged BTZ BH. Unlike the latter however, these electrically charged HBTZ BHs are not given in closed form but are
   constructed numerically. We have found it useful to consider some thermodynamic properties of the electrically charged HBTZ BHs, by way
   of contrasting them to the electrically charged BTZ BHs.

   As this is a preliminary exploration of such systems, there is a long list of follow-up investigations, which we list:

   \begin{itemize}
   \item
     Seek torsionful solutions in the model studied here.
   \item
     Analyze the present model, augmented by the planar Skyrme model in $2+1$ dimensions, by way of introducing a metter field.
     This would be an alternative to the Maxwell field considered here as a matter field. We expect that with the Skyrme matter,
     regular solutions in the limit of vanishing horizon radius may exist.
   \item
     Consider the HBTZ BHs in $2+1$ dimensions, of the next HCSG model in the hierarchy, namely \re{HCS38} that results from
     the dimensional reduction of the $4$-th CP density in $8$ dimensions. We expect the resulting HBTZ BH to be qualitaively similar
     to the one studied here.
   \item
     Consider the HBTZ BHs in $4+1$ and $6+1$ dimensions, in the HCSG models constructed from HCS densities extracted from HCP densities
     descended from the $4$-th and $5$-th CP densities given in Ref.~\cite{Tchrakian:2017fdw}.

     This would throw light on some general properties of HCSG BHs in all odd dimensions, analogous to the property common to CSG BHs
     in all odd dimensions, namely that the simplest solutions posesses a generic form with $\frac{1}{g_{rr}}=-g_{tt}=\ka^2r^2+c_t$,
     starting with the BTZ BH in $2+1$ dimensions.
   \item
     Consider the HBTZ BHs in $3+1$ dimensions, in the HCSG models constructed from HCS densities extracted from HCP densities
     descended from the $3$-rd and $4$-th CP densities~~\cite{Tchrakian:2017fdw}. This would be a novel result in that Chern-Simons
     gravitational BHs in {\it even} dimensional spacetimes may also be of some interest.
   \end{itemize}
   
   In addition to the above systematic follow-ups, it is possible to consider a more formal possibility. This is the analogue of augmenting
   a given gravitational Lagrangian, with a gravitational Chern-Simons (GCS) term, most familiary known since a long time from
   Ref.~\cite{Deser:1982vy} where the usual Einstein Lagrangian was augmented by the GCS term in $2+1$ dimensions. In the present case
   for example, one might consider the gravitational Higgs--Chern-Simons term (GHCS), since this one features the same field multiplets.
   $(e_\mu^a,\f^a,\f)$ as the HCSG model studied. (The definitions of GCHS terms, along with the usual GCS densities, is given in
   Ref.~\cite{Tchrakian:2017fdw}. These are derived from a {\it modified version}~\footnote{In odd dimensions, where the HCS term is
   extracted from a HCP term in even dimensions, the chiral matrix under the Trace of ther latter must be removed {\it by hand}. This
   {\it ad hoc} step is taken so that the leading term in the GHCS density is the GCS density in the given dimension. This question does not
   arise in even dimensional spacetimes.
   .} of the HCS densities rather than from the CS densities in the $usual$ case.)
 
The simplest such GHCS term in $2+1$ dimensions, which has the same dimensions as \re{HCSG2+1}, is
\bea
\hat\Om_{\rm GHCS}^{(3,6)}&=&-2\eta^2\hat\Om_{\rm GCS}^{{(3)}}+4\vep^{\la\mu\nu}\,\f^a R_{\mu\nu}^{ab}D_\la\f^b,
\label{GHCS36}
\\
\hat\Om_{\rm GCS}^{{(3)}}&=&-\frac{1}{2}\,\vep^{\la\mu\nu}
\om_{\la}^{ab}\left[R_{\mu\nu}^{ab}-\frac23\left(\om_{\mu}\om_{\nu}\right)^{ab}\right],
\label{GCS31}
\eea
$\hat\Om_{\rm GCS}^{{(3)}}$ in \re{GCS31} being the usual GCS term used in Ref.~\cite{Deser:1982vy,Deser:1981wh}.                     
Note that the GHCS density \re{GHCS36} features only the frame-vector field $\f^a$ and not the scalar $\f=\f^4$, while the latter
is present in the HCSG model \re{HCSG2+1}.

Perhaps more interesting are the first two GHCS densities in $3+1$ dimensions, each derived from the HCS
densities extracted, respectively, from HCP density densities in $6$ and in $8$ dimensions. These are
\bea
\hat\Om_{\rm GHCS}^{(4,6)}&=&
-\frac14\,\vep^{\mu\nu\rho\si}\,R_{\mu\nu}^{ab}\,R_{\rho\si}^{ab}\,\f\label{GHCS46}\\
\hat\Om_{\rm GHCS}^{(4,8)}&=&
-\vep^{\mu\nu\rho\si}\,
R_{\mu\nu}^{ab}\left\{\left[\frac18\left(1-\frac{1}{3}(|\f^a|^2+\f^2)\right)R_{\rho\si}^{ab}
+\frac13\,\f_{\rho\si}^{ab}\right]\f+\frac43\f^{a}D_{\rho}\f^{b}\,\pa_{\si}\f\right\}\label{GHCS48}
\eea
where the abbreviated notation
\[
\f_{\mu\nu}^{ab}=D_{[\mu}\f^a D_{\nu]}\f^b\ , \quad a=1,2,3,4\ ;\quad{\rm and}\quad \f=\f^5\,.
\]                             
                              The reason for displaying \re{GHCS48} in addition to \re{GHCS46} is that the former does not feature any
                              dynamical terms for $(\f^a,\f)$, while the latter does. The GHCS densities \re{GHCS46} and \re{GHCS48}
                              can be viewed as alternatives to the GHCS density proposed in Ref.~\cite{Jackiw:2003pm}.

 \medskip
\noindent {\large\bf Acknowledgements}
\\
We are grateful to Ruben Manvelyan for helpful discussions throughout this work, and to Hermann Nicolai for hospitality at the
Albert-Einstein Institute for Gravitational physics (Golm), where this work was started. 
Thanks to 
Brian Dolan, 
Carlos Herdeiro, 
Olaf Lechtenfeld
and Shahin Sheikh-Jabbari for helpful comments. 
E.R. acknowledges funding from the FCT-IF programme and is 
also partially supported by the  H2020-MSCA-RISE-2015 Grant No.  StronGrHEP-690904, 
the H2020-MSCA-RISE-2017 Grant No. FunFiCO-777740,
and by the CIDMA project UID/MAT/04106/2013. 
 
\begin{small}

\end{small}


\begin{thebibliography}{99}
\bibitem{Witten:1988hc}
  E.~Witten,
  ``(2+1)-dimensional gravity as an exactly soluble system,''
  Nucl.\ Phys.\ B {\bf 311} (1988) 46.
\bibitem{Chamseddine:1989nu}
  A.~H.~Chamseddine,
  ``Topological gauge theory of gravity in five-dimensions and all odd dimensions,''
  Phys.\ Lett.\ B {\bf 233} (1989) 291.
\bibitem{Chamseddine:1990gk}
  A.~H.~Chamseddine,
  ``Topological gravity and supergravity in various dimensions,''
  Nucl.\ Phys.\ B {\bf 346} (1990) 213.
\bibitem{Tchrakian:2010ar}
  D.~H.~Tchrakian,
  ``Notes on Yang-Mills--Higgs monopoles and dyons on $\R^D$, and Chern-Simons--Higgs solitons on $\R^{D-2}$: Dimensional reduction of Chern-Pontryagin densities,''
  J.\ Phys.\ A {\bf 44} (2011) 343001
  [arXiv:1009.3790 [hep-th]].
\bibitem{Radu:2011zy}
  E.~Radu and D.~H.~Tchrakian,
  ``New Chern-Simons densities in both odd and even dimensions,''
  arXiv:1101.5068 [hep-th].
\bibitem{Tchrakian:2015pka}
  D.~H.~Tchrakian,
  ``Higgs-and Skyrme-Chern-Simons densities in all dimensions,''
  J.\ Phys.\ A {\bf 48} (2015) no.37,  375401
  [arXiv:1505.05344 [hep-th]].
 \bibitem{Szabo:2014zua}
   R.~J.~Szabo and O.~Valdivia,
  ``Covariant quiver gauge theories,''
   JHEP {\bf 1406} (2014) 144,
   [arXiv:1404.4319 [hep-th]].
\bibitem{Tchrakian:2017fdw}
  D.~H.~Tchrakian,
  ``Chern-Simons gravities (CSG) and gravitational Chern-Simons (GCS) densities in all dimensions,''
  arXiv:1712.05190 [gr-qc].
\bibitem{Banados:1992wn}
  M.~Banados, C.~Teitelboim and J.~Zanelli,
  ``The black hole in three-dimensional space-time,''
  Phys.\ Rev.\ Lett.\  {\bf 69} (1992) 1849
  [hep-th/9204099].
\bibitem{Gubser:2008px}
  S.~S.~Gubser,
  ``Breaking an Abelian gauge symmetry near a black hole horizon,''
  Phys.\ Rev.\ D {\bf 78} (2008) 065034
  [arXiv:0801.2977 [hep-th]].
\bibitem{Gubser:2008zu}
  S.~S.~Gubser,
  ``Colorful horizons with charge in anti-de Sitter space,''
  Phys.\ Rev.\ Lett.\  {\bf 101} (2008) 191601
  [arXiv:0803.3483 [hep-th]].
\bibitem{Winstanley:1998sn}
  E.~Winstanley,
  ``Existence of stable hairy black holes in SU(2) Einstein-Yang-Mills theory with a negative cosmological constant,''
  Class.\ Quant.\ Grav.\  {\bf 16} (1999) 1963
  [gr-qc/9812064].
\bibitem{Charap:1977ww}
  J.~M.~Charap and M.~J.~Duff,
  ``Gravitational effects on Yang-Mills topology,''
  Phys.\ Lett.\  {\bf 69B} (1977) 445.
\bibitem{Brihaye:2006bk}
  Y.~Brihaye and E.~Radu,
  ``On d=4 Yang-Mills instantons in a spherically symmetric background,''
  Europhys.\ Lett.\  {\bf 75} (2006) 730
  [hep-th/0605111].
\bibitem{Radu:2007az}
  E.~Radu, D.~H.~Tchrakian and Y.~Yang,
  ``Spherically symmetric selfdual Yang-Mills instantons on curved backgrounds in all even dimensions,''
  Phys.\ Rev.\ D {\bf 77} (2008) 044017
  [arXiv:0707.1270 [hep-th]].
\bibitem{Bekenstein:1996pn}
  J.~D.~Bekenstein,
  ``Black hole hair: 25 - years after,''
  in "Moscow 1996, 2nd International A.D. Sakharov Conference on physics", pp. 216-219
  [gr-qc/9605059].
\bibitem{Volkov:1998cc}
  M.~S.~Volkov and D.~V.~Gal'tsov,
  ``Gravitating non-Abelian solitons and black holes with Yang-Mills fields,''
  Phys.\ Rept.\  {\bf 319} (1999) 1
  [hep-th/9810070].
\bibitem{Herdeiro:2015waa}
  C.~A.~R.~Herdeiro and E.~Radu,
  ``Asymptotically flat black holes with scalar hair: a review,''
  Int.\ J.\ Mod.\ Phys.\ D {\bf 24} (2015) no.09,  1542014
  [arXiv:1504.08209 [gr-qc]].
\bibitem{Deser:1982vy}
  S.~Deser, R.~Jackiw and S.~Templeton,
  ``Three-dimensional massive gauge theories,''
  Phys.\ Rev.\ Lett.\  {\bf 48} (1982) 975.
\bibitem{Deser:1981wh}
  S.~Deser, R.~Jackiw and S.~Templeton,
  ``Topologically massive gauge theories,''
  Annals Phys.\  {\bf 140} (1982) 372
   [Annals Phys.\  {\bf 281} (2000) 409]
   Erratum: [Annals Phys.\  {\bf 185} (1988) 406].
\bibitem{Jackiw:2003pm}
  R.~Jackiw and S.~Y.~Pi,
  ``Chern-Simons modification of general relativity,''
  Phys.\ Rev.\ D {\bf 68} (2003) 104012
  [gr-qc/0308071].
\end{thebibliography}
\end{document}